\documentstyle[prl,aps,epsfig,multicol]{revtex}
\begin{document}

\title{Optical decay from a {Fabry-Perot} cavity \\
faster than the decay time}
\author{H. Rohde, J. Eschner, F. Schmidt-Kaler, R. Blatt}
\address{Experimentalphysik, Univ. Innsbruck, A-6020 Austria}
\maketitle

\begin{abstract}
The dynamical response of an optical {Fabry-Perot} cavity is
investigated experimentally. We observe oscillations in the
transmitted and reflected light intensity if the frequency of the
incoupled light field is rapidly changed. In addition, the decay
of a cavity-stored light field is accelerated if the phase and
intensity of the incoupled light are switched in an appropriate
way. The theoretical model by M. J. Lawrence {\em et al}, JOSA B
16, 523 (1999) agrees with our observations.

\end{abstract}


\begin{multicols}{2}

\section{Introduction}
Optical {Fabry-Perot} cavities are common in laser spectroscopy
and interferometry and are frequently used for the stabilization
of laser sources \cite{BERGQUIST00,BERGQUIST99}. A novel
application for high-finesse cavities was proposed in the early
90's in the upcoming field of cavity-QED: Single atoms are
strongly coupled to a cavity-stored photon field such that the
mutual coherent oscillatory exchange between both sub-systems is
much faster than their individual decay rates. A number of
experiments, in both the optical \cite{TURCHETTE95} and in the
microwave regime \cite{RAUSCHENBEUTEL99}, have shown that this
cavity-atom coupling can indeed be used for quantum information
processing \cite{ALMAGRO}, where single quantum systems such as
atoms or photons carry qubits (as the quantum alternative for the
well known bits in information science).

We are interested in the physics of Fabry-Perot cavities because
they can be used as an interface between static quantum
information storage, which is advantageously done in long lived
states of trapped atoms, and the transport of quantum information
by light \cite{BRIEGEL98}. For this we are currently investigating
the physical system composed of a single trapped $^{40}Ca^+$ ion
inside the waist of a high finesse cavity \cite{SFB}.

A second major motivation is the use of light forces on atoms in
cavities which leads to trapping and cooling, as published
recently \cite{PINSKE00,KIMBLE00}. Here, light fields at the
single photon level lead to significant forces due to the large
enhancement factor of a high finesse cavity. The next generation
of experiments will detect the atom's position \cite{MABUCHI99}
and even modify the light field in order to stabilize the atom's
trajectory. This, however, not only demands a fast detection
method, but also a technique to rapidly alter the photon field
inside the optical Fabry-Perot cavity.

In this paper we show how the input light can be modulated to
observe enforced evacuation of a cavity-stored light field, within
a timescale below the cavity damping time. For this, the phase and
intensity of the incoupled light field are switched rapidly such
that destructive interference nulls the cavity-stored field. This
technique is obviously interesting for application in the above
experiments.

The dynamic response of a Fabry-Perot cabity has been investigated
in the context of the Laser Interferometer Gravitational-Wave
Observatory (LIGO) by M. J. Lawrence {\em et al.}
\cite{LAWRENCE99}. In that publication the transmitted light
intensity and the Pound-Drever-Hall (PDH) error signal in
reflection \cite{PDH} are studied when the cavity length or the
frequency of the input light field is rapidly changed. The results
which we present here complete and generalize those earlier
studies since we show how the full control of frequency, phase,
intensity and time duration of an input light field leads to a
much larger variety of interference effects between cavity-stored
light and input field.

The paper is organized as follows: After describing the
experimental setup, we first study the response of the transmitted
and reflected light when the input field frequency is swept
continuously over the cavity resonance. Our results fully recover
and confirm the findings of M. J. Lawrence {\em et al.}
\cite{LAWRENCE99}. In the following parts we describe more
sophisticated implementations of modulating the input, such as
frequency switching (covered by section 3) and phase switching
(section 4). In the final section we present the unusual response
of the cavity field to simultaneous phase and intensity switching,
whereby the decay of the cavity field, as measured in the
transmitted light, is accelerated much below the cavity decay
time.

\section{Experimental setup}
A Titanium-Sapphire laser delivers up to 500~mW of light at 729~nm
wavelength. One part of the laser power is focused into a Paul
trap and used to excite the narrow $S_{1/2} \leftrightarrow
D_{5/2}$ transition of single trapped $^{40}Ca^+$ ions
\cite{FERDI00}. A second part is used to generate a PDH error
signal from the reflection of a reference cavity with finesse
220000 (mirrors optically contacted to a 20~cm long ULE-spacer).
The Titanium-Sapphire laser frequency is stabilized to this
reference cavity and we measure a laser linewidth of less than
100~Hz (using the narrow $Ca^+$ transition) \cite{Laserlinewidth}.

A third fraction of the light is coupled into an optical fiber
which transports it from the optical table to a vibration isolated
platform. The experimental setup on that platform is depicted in
Fig.~1. The light from the fiber output is sent through an
acousto-optic modulator (AOM: Brimrose Inc. USA, GRP-650) in
double-pass such that the light in the first order beam is
frequency shifted by twice the AOM driving frequency (which can be
varied from 550 to 750~MHz). After passing through an
electro-optical modulator (EOM: Linos Inc. Germany, LM0202) to add
sidebands at 17~MHz to the light, the light is mode-matched into a
second high finesse Fabry-Perot cavity. This cavity consists of a
ULE-spacer of 15~cm length with a central bore and two high
reflecting mirrors (Research Electro-Optics Inc. USA´) optically
contacted onto the polished front faces of the spacer. The cavity
rests on 4 pins inside a UHV-vessel at $\le 10^{-8}$~mbar. The
UHV-vessel is temperature stabilized to 25~$\pm$~0.05~C. Light
transmitted by the cavity is monitored on a photo-diode PD~1
(bandwidth $\approx$ 0.1~MHz) and a CCD-camera. The reflected
intensity is detected by a photo-diode PD~2 (bandwidth $\approx$
100~MHz), which is used for the generation of a PDH error signal.

\begin{center}
\begin{figure}[tbp]
\epsfxsize=0.85\hsize \epsfbox{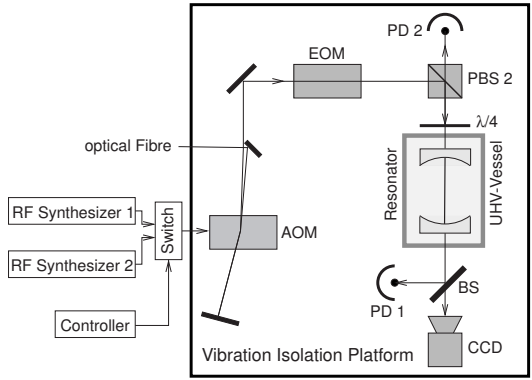} \vspace{\baselineskip}
\caption{Optical setup of the high-finesse Fabry-Perot cavity on
the isolation platform (PBS: polarizing beam splitter, BS: beam
splitter, PD: photo-diode, $\lambda$/2: half-wave plate, AOM:
acousto-optic modulator, EOM: electro-optic modulator, see text
for further details). \label{fig1}}
\end{figure}
\end{center}

An RF-synthesizer provides the driving frequency for the AOM. The
light frequency is varied over the TEM$_{00}$ resonance of the
cavity. A second synthesizer of the same type and synchronized to
the same timebase can be set to a well-defined frequency and phase
offset relative to the first one. An RF-switch (Minicircuits
ZYSWA-2-50, switching time 3~ns) is used to switch rapidly between
the first and the second RF-synthesizer, in order to generate
controlled frequency, intensity or phase jumps of the input light.

\begin{center}
\begin{figure}[tbp]
\epsfxsize=0.9\hsize \epsfbox{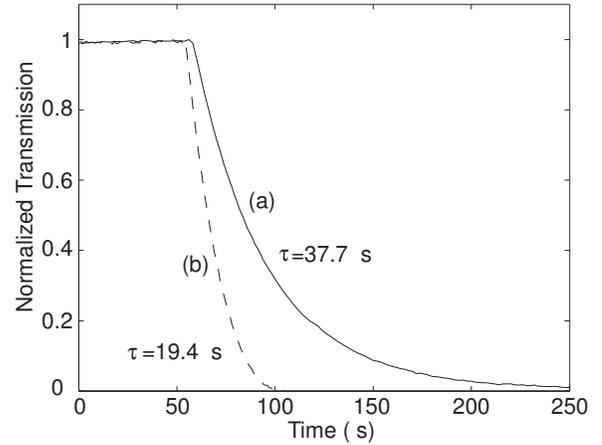} \caption {a) Decay of a
stored cavity field as monitored in transmission (solid line). b)
The cavity is filled with a resonant light field. We observe
accelerated cavity decay (broken line) monitored in transmission,
when the phase of the input light field is switched by 180~degree
(near $t$= 50~$\mu$s). Finally, the input light is switched off
(near $t$= 100~$\mu$s). The ratio of both measured decay constants
of 1.95(0.05) agrees with the expected factor two. \label{fig2}}
\end{figure}
\end{center}

First, we measured the cavity decay time: The RF-drive of the AOM
is switched off when the cavity transmission is near a maximum
(TEM$_{00}$ resonance) and the exponential decay of the
cavity-stored light field is observed with PD~1 and an
oscilloscope, see Fig.~\ref{fig2}~a. We find an exponential decay
with a time constant of $\tau_s$=37.1(0.3)~$\mu$s. The finesse $F$
of the cavity is related to decay time $\tau_s=FL/c\pi$, which
allows us to determine $F$ to be 233000~$\pm$~2000.

\section{Input light frequency modulation}
Consider a linear sweep of the input light frequency and a
standing wave cavity of fixed length $L$. Initially at $t=0$, the
input field with constant amplitude $E_0$ is at frequency
$\omega_0$, and before the first round-trip the intra-cavity field
reads $E_{cav}^{(0)} = i\sqrt{T}E_0 e^{-i\omega_0 t}$. Here, $T$
denotes the transmission of the input coupling mirror. After one
round-trip, the cavity field is slightly attenuated, which the
round-trip loss $\rho = 1-\pi/F$ accounts for, and interferes with
the transmitted input field, which is now at different frequency
$\omega_{\tau}$. We obtain $E_{cav}^{(1)} = \rho e^{-i\omega_0
(t-\tau)} + i\sqrt{T}E_0 e^{-i\omega_{\tau} \tau}
e^{-i\omega_{\tau} (t-\tau))}.$ For $n$ round-trips, a recursion
equation is found \cite{LAWRENCE99} for $E_{cav}^{(n)}$, which
finally leads to the differential equation
\begin{center}
\begin{equation}
E_{cav}/d\tilde{t} = -(1 - i \nu_{\omega} \tilde{t}) E_{cav} + i
(\sqrt{T} F/\pi) E_0,
\end{equation}
\end{center}
with $\tilde{t}=t/\tau_s$, the cavity decay time $\tau_s$ defined
as above, and $\nu_{\omega}= 2 FL \dot{\omega}\tau/\pi c$ which
denotes the normalized scan rate. As discussed in reference
\cite{LAWRENCE99}, the cavity transmission signal and the PDH
error signals exhibit oscillations, which appear for
$\nu_{\omega} \geq 1$.

\end{multicols}

\begin{center}
\begin{figure}[tbp]
\epsfxsize=0.8\hsize \epsfbox{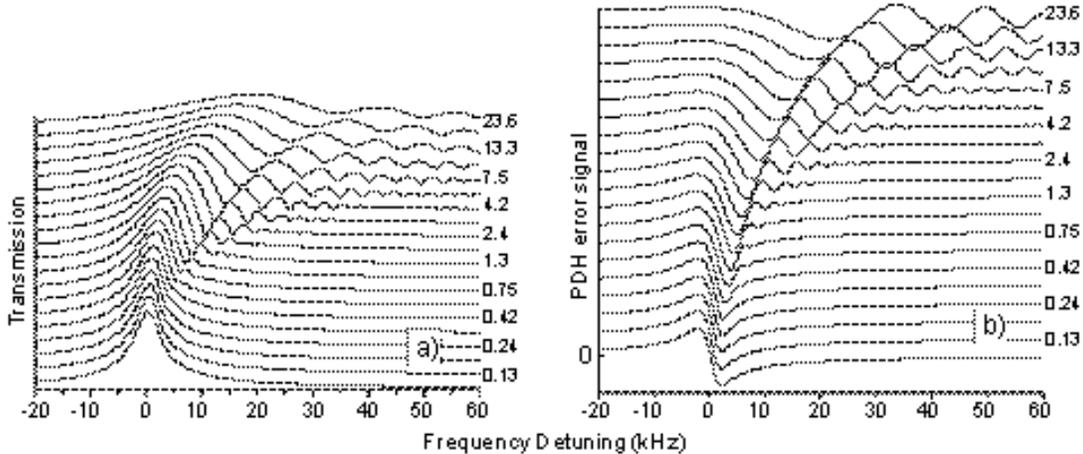} \caption{a) Simulation of
the cavity intensity transmission, when the input frequency is
varied over the resonance. We plot $|E_{cav}|^2$ of Eq.1. For this
picture, the scan rate $\nu_{\omega}$ is chosen 0.1 for the lowest
trace and increased by $133\%$ each step in 20 steps, as indicated
at the right hand side of the plot (for clarity, the curves are
shifted upwards with increasing $\nu_{\omega})$). As the
calculation shows, for high $\nu_{\omega}$, the point of highest
transmission is shifted towards higher optical frequency detuning
and the transmission level is lowered. b) Simulated PDH error
signal, $Re(E_{cav})$, Eq.~8 in reference~[11] for different
values of $\nu_{\omega}$.} \label{fig3}
\end{figure}
\end{center}

\begin{multicols}{2}

\begin{center}
\begin{figure}[tbp]
\epsfxsize=0.8\hsize \epsfbox{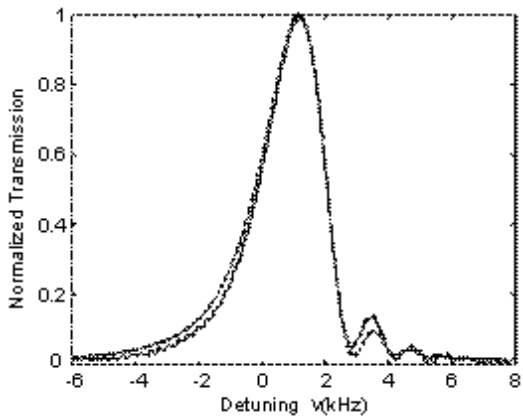} \caption{Cavity
transmission as measured (solid line) and calculated (broken) for
$\nu_{\omega}=1.35$.\label{fig4}}
\end{figure}
\end{center}

Fig.~3 shows a numeric solution of Eq.~1, for different
$\nu_{\omega}$. For the simulation, we take the values which are
realized in the experiment for a standing wave cavity with
$F$=233000 and  length of $L$=0.15~m. The frequency axis is
calculated using the transformation from time steps of $\tau_s$,
as used for the evaluation of the differential equation, to the
frequency detuning in $Hz$, by multiplying  $\tau_s$ with
$\nu_{\omega} \cdot c /(4 F L)$. The experimental result for a
cavity transmission signal is shown in Fig.~4, together with the
simulation. The frequency scan rate was $\dot{\omega}= 2\pi $
64~MHz/s which results in a $\nu_{\omega}$ of 1.27. Fig.~5 shows
PDH error signals, as obtained from the de-modulated signal of
PD2. Again, for values of $\nu_{\omega}$ approaching one, the
error signal is distorted, and shows oscillations.

\begin{center}
\begin{figure}[tbp]
\epsfxsize=0.85\hsize \epsfbox{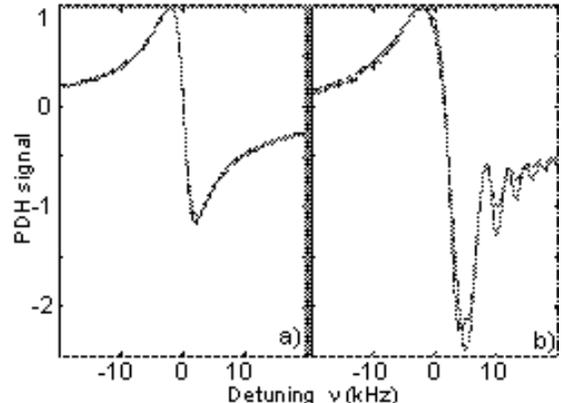} \caption{PDH error signal
as measured for different $\nu_{\omega}$ of 0.01, and 1.35
(experiment: solid line, simulation: dotted line). \label{fig5}}
\end{figure}
\end{center}

\section{Switching the input light frequency}
In another experiment, we used two RF-synthesizers to switch the
frequency of the input light field. First the input is kept in
resonance with the cavity and the cavity is filled. Then the
frequency of the input light is switched by 46~kHz and the output
light intensity behind the cavity is monitored. We observe an
exponential decay, which is sinusoidally modulated at the beat
frequency between the two input fields, see Fig.~6 for the data.
We observe that the initial phase of this modulation is changed
when the phase between both synthesizers varies.

\section{Phase modulation response}
As in the previous section, we use two synthesizers to generate
the drive for the AOM. Both frequencies are kept in resonance with
the fundamental cavity  mode. The cavity is filled with the first
source, then we switch to the second source whose phase is at
180~degree to the first one. The intensity behind the cavity is
monitored. Our result is shown in Fig.~2b, where we observe an
exponential decay and fit a decay constant of 19.4~(0.4)$\mu$s.
Note that this decay is a factor of 1.95(0.05) faster than the
free cavity decay, observed after simply cutting off the input
light (Fig.~2a). This experimental finding agrees well with the
expected factor of two, since the input field intensity was kept
constant. The second light field, with opposite phase to the first
one, builds up inside the cavity and leads there to destructive
interference. At the instant, when the transmitted light from the
cavity vanishes, we cut the second input field to leave the cavity
empty (Fig.2~b, $t \simeq 100~\mu$s). One may wonder how the
stored light energy escapes: It leaves the cavity via the input
coupling mirror, where it interferes constructivly with the input
light field. These findings lead immediately to the idea that the
decay time of light in the cavity can be shortened even beyond a
factor of two if phase {\em and} intensity of the second input
field are chosen properly.

\begin{center}
\begin{figure}[tbp]
\epsfxsize=0.85\hsize \epsfbox{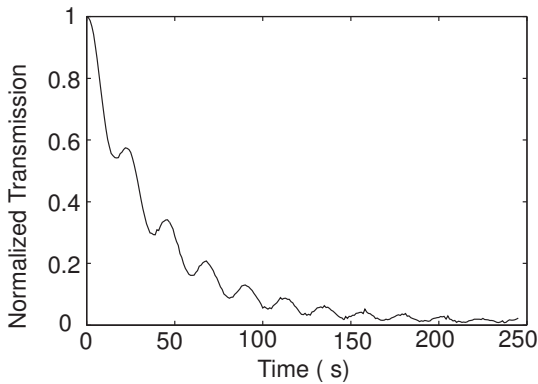} \caption{The cavity is
filled at resonance, then at $t=0$ the input frequency is switched
by 46~kHz. \label{fig6}}
\end{figure}
\end{center}

\section{Phase and intensity modulation and accelerated cavity decay}
As in the previous section, two input fields are used. The first,
on resonance, fills the cavity and the second is used to induce a
cavity decay as fast as possible. We set the intensity to twice
that of the first input field, and indeed observed a
$\sim$~2.3-times faster decay, with 16.9(0.7)~$\mu$s. However, for
the intensity switching, the AOM drive power is increased, which
also transiently affects the phase. Thus, with the used setup it
was impossible to adjust the second light field's phase to be
exactly opposite to the first one and to achieve an even faster
cancellation of the cavity field.

\section{Conclusion}
We have shown experimentally interference effects which occur when
the input light field of an optical cavity is rapidly altered in
frequency, phase and intensity. The transmitted intensity is
monitored to reveal a strongly modified cavity decay. In addition,
the Pound-Drever-Hall error signal in reflection, as used for
laser frequency stabilization, shows oscillations when the laser
frequency is scanned quickly over the cavity resonance. The PDH
signal, usually an antisymmetric signal with a zero crossing
(locking point) exactly at cavity resonance, may then even exhibit
a few oscillations and multiple zero crossings (for $\nu_{\omega}
\geq 1$), and loses completely its shape. The consequence is
clear: If an ultra high finesse cavity (with narrow linewidth) is
used for the frequency stabilization of a laser with high
frequency fluctuations, the servo-loop for the frequency locking
will initially have some problems to stabilize the laser's
frequency within the reference cavity's linewidth. Thus, a
pre-stabilization \cite{BERGQUIST99} of the laser becomes
necessary to reduce its initial frequency jitter.

We have shown how the cavity decay time can be decreased below its
normal value. This could have future applications if a stored
cavity field has to be changed rapidly. All observations agree
well with the theoretical predictions, which are based on the
interference of electromagnetic fields, and which are outlined in
this paper and in  M. J. Lawrence {\em et al} \cite{LAWRENCE99}.

This work is supported by the SFB-15 of the Austrian "Fonds zur
F\"orderung der wissenschaftliche Forschung", the European
Commission (TMR networks QI and QUEST) and the "Institut f\"ur
Quanteninformation GmbH".

\end{multicols}

\newpage


\begin{thebibliography}{}

\bibitem{BERGQUIST00}
R. J. Rafac et al,  {Sub-decahertz ultraviolet spectroscopy of
$^{199}${Hg}$^+$}, Phys. Rev. Lett. {\bf 85}, 2462 (2000).

\bibitem{BERGQUIST99}
B. C. Young et al,  {Visible lasers with subhertz linewidths},
Phys. Rev. Lett. {\bf 82}, 3799 (1999).

\bibitem{TURCHETTE95}
Q.~Turchette et al, {Measurement of conditional phase shifts for
quantum logic}, Phys. Rev. Lett. {\bf 75}, 4710 (1995).

\bibitem{RAUSCHENBEUTEL99}
A.~Rauschenbeutel et al, {Coherent operation of a tunable quantum
phase gate in cavity-QED}, Phys. Rev. Lett. {\bf 83}, 5166 (1999).

\bibitem{ALMAGRO}
{The Physics of Quantum Information}, Springer, Berin. ed. D.
Bouwmeester, A. Ekert, and A. Zeilinger (2000).

\bibitem{BRIEGEL98}
H.-J. Briegel et al, {Quantum repeaters: The role of imperfect
local operations in  quantum communication}, Phys. Rev. Lett. {\bf
81}, 5932 (1998).

\bibitem{SFB}
{Application for the Austian Sience Foundation {SFB12-P2}: Control
and measurement of coherent systems}, Innsbruck.

\bibitem{PINSKE00}
P.~W. Pinske, T.~Fischer, P.~Maunz, and G.~Rempe, {Trapping an
atom with single photons}, Nature {\bf 404}, 365 (2000).

\bibitem{KIMBLE00}
C.~Hood et al, {The atom cavity microscope - single atoms bound in
orbit by single photons}, Science {\bf 287}, 1447 (2000).

\bibitem{MABUCHI99} H. Mabuchi, J. Ye, and H. J. Kimble,
{Full observation of single-atom dynamics in cavity-QED}, Appl.
Phys. B68, 1095 (1999).

\bibitem{LAWRENCE99}
M. J. Lawrence et al, {Dynamic response of a Fabry-Perot
interferometer}, JOSA B 16, 523 (1999)

\bibitem{PDH}
R. W. P. Drever et al, {Laser Phase and Frequency Stabilization
using an optical Resonator}, Appl. Phys B31, 97 (1983), A.
Schenzle, R. DeVoe, and R. Brewer, {Phase-modulation lase
spectroscopy}, Phys. Rev. A25, 2606 (1982)

\bibitem{FERDI00}
F. Schmidt-Kaler et al, {Ground state cooling, quantum state
engineering and study of decoherence of ions in paul traps}, J.
Mod. Opt. {\bf 47}, 2573 (2000)

\bibitem{Laserlinewidth}
H. Rohde, PhD-thesis, Innsbruck 2000, unpublished


\end{thebibliography}
\end{document}